\documentclass[aps,prd,floats,preprintnumbers,nofootinbib]{revtex4}
\usepackage{epsfig}

\begin{document}

\preprint{NUHEP-TH/08-05}

\title{Light Sterile Neutrino Effects at $\theta_{13}$-Sensitive Reactor Neutrino Experiments}

\author{Andr\'e de Gouv\^ea}
\affiliation{Northwestern University, Department of Physics \& Astronomy, 2145 Sheridan Road, Evanston, IL~60208, USA}

\author{Thomas Wytock}
\affiliation{Northwestern University, Department of Physics \& Astronomy, 2145 Sheridan Road, Evanston, IL~60208, USA}


\begin{abstract}

We study the impact of very light sterile neutrinos ($\Delta m^2_{\rm new}\in[1,10]\times 10^{-2}$~eV$^2$, $\sin^22\theta_{\rm new}<10^{-1}$) on upcoming $\theta_{13}$-driven reactor antineutrino experiments like Double-CHOOZ and Daya Bay. Oscillations driven by these vales of $\Delta m^2_{\rm new}$ affect data in the near and far detectors differently and hence potentially modify the capability of these experimental setups to constrain and measure $\sin^22\theta_{13}$. We find that the hypothesis $\theta_{\rm new}\neq 0$ negatively impacts one's ability to either place an upper bound on $\sin^22\theta_{13}$ in the advent of no oscillation signal or measure $\sin^22\theta_{13}$ if a $\theta_{13}$-driven signal is observed. The impact of sterile neutrino effects, however, depends significantly on one's ability to measure the recoil positron energy spectrum. If $\sin^22\theta_{\rm new}\gtrsim 10^{-2}$, upcoming $\theta_{13}$-driven reactor antineutrino experiments should be able to measure $\sin^22\theta_{\rm new}$ and $\Delta m^2_{\rm new}$, along with $\sin^22\theta_{13}$,  as long as one is sensitive to distortions in the recoil positron energy spectrum in the near (and far) detectors.

\end{abstract}

\maketitle

\setcounter{equation}{0}
\section{Introduction}

Next-generation reactor antineutrino experiments are currently under construction (for details on concrete projects see \cite{Ardellier:2006mn,Guo:2007ug}). Guided by the results of current neutrino oscillation experiments \cite{TASI}, these are aimed at observing electron antineutrino disappearance driven by the ``atmospheric'' mass-squared difference, $\Delta m^2_{13}$, and hence measuring the still elusive $\theta_{13}$ mixing angle ({\it cf.} Sec.~\ref{sec:mixing}).

In order to significantly improve on previous reactor antineutrino experiments, next-generation experiments will make use of a two-detector setup. A far detector is to be placed an optimal distance away from the reactor core so as to maximize $\theta_{13}$-driven electron antineutrino disappearance, while a near detector is placed close to the reactor core in order to measure the ``unoscillated'' electron antineutrino flux. Assuming differences between near and far detectors are understood at the few permille level and that enough statistics are accumulated, one ultimately aims at being sensitive to $\sin^22\theta_{13}$ values  around $1\%$. 

The setup summarized above relies on the assumption that all neutrino oscillation phenomena are properly described in terms of three massive neutrinos which interact under the well-known weak interactions. Current data reveal this to be an excellent assumption, but the possibility of other ``subleading'' effects remains. Here, we consider the possibility that light sterile neutrinos mix slightly with active neutrinos and hence modify the standard neutrino oscillation picture. 

Sterile neutrinos --- gauge singlet fermions --- are among the simplest extensions of the standard model of particle physics. They may be an integral part of the physics responsible for neutrino masses \cite{seesaw,lowseesaw}, and may be a component of the dark matter in the Universe \cite{dm_seesaw}. Theoretically speaking, nothing is known about sterile neutrino masses, and very light sterile neutrinos (say $m_{\nu}\ll 1$~keV) are as natural (as defined by 'tHooft) as very heavy ones (say $m_{\nu}\gg 10^{10}$~GeV). Experiments, therefore, provide almost all the unbiased information we have regarding sterile neutrino masses. We briefly discuss experimental constraints in Sec.~\ref{sec:mixing}.

We concentrate on sterile neutrinos that could qualitatively affect the interpretation of $\theta_{13}$-driven reactor neutrino experiments. We find that this can happen for very light sterile neutrinos that introduce to the three-neutrino-oscillation picture a new mass-squared difference of order $(1-10)\times10^{-2}$~eV$^2$. For example, the ``hypothetical presence'' of such sterile states hinders the ability of these experiments to rule out certain values of $\theta_{13}$ if no evidence for oscillations is observed. On the other hand, small sterile neutrino effects, if present, may lead to very significant oscillatory effects in the near detector even in the limit of vanishing $\theta_{13}$. 
Detailed results are presented in Sec.~\ref{sec:numerical}, along with a study of how well $\theta_{13}$ and sterile mixing parameters can be simultaneously measured with next-generation reactor neutrino data. 

Before proceeding, we'd like to highlight our goals and the limitations of our analysis. We are interested in discussing the fact that sterile neutrinos can significantly alter the interpretation of the comparison between near and far detector data in reactor antineutrino experiments. 
We would also like to point out that the near detector --- because it is expected to collect hundreds of thousands of neutrino scattering events --- may play an active role in revealing new, unexpected physics. With this in mind, our simulations, discussed in more detail in Sec.~\ref{sec:numerical}, are not aimed at realistically describing experimental setups or quantitatively gauging their capabilities. Qualitatively, however, we believe that all of the effects discussed in this manuscript will manifest themselves once real experimental data is analyzed. Our main message, along with our results, is summarized in Sec.~\ref{sec:conclusion}.

\setcounter{equation}{0}
\section{Very Light Sterile Neutrino Effects at Reactor Experiments}
\label{sec:mixing}

We are adding to the three active neutrinos a fourth sterile state and are hence faced with four neutrino mass eigenstates. The electron neutrino $\nu_e$ can be expressed as a linear combination of $\nu_i$, $i=1,2,3,4$: $\nu_e=\sum_iU_{ei}\nu_i$. We order the neutrino masses as follows (for a detailed discussion see \cite{deGouvea:2008nm}). $\nu_4$ is the ``mostly sterile'' state ($|U_{s4}|^2\gg |U_{s1,s2,s3}|^2$) while $\nu_{1,2,3}$ are mostly active. $\nu_{1,2,3}$ are defined in the ``usual way'': $m_1^2<m_2^2$ while $|m_3^2-m_1^2|>m^2_2-m_1^2$ and $|m_3^2-m_2^2|>m_2^2-m_1^2$. The case $m_3^2>m_2^2$ is referred to as the `normal' mass hierarchy, while the case $m_3^2<m_1^2$ is referred to as the `inverted' mass hierarchy. See \cite{TASI,deGouvea:2008nm} for more details.

The electron neutrino (or antineutrino) survival probability is
\begin{eqnarray}
P_{ee}&=&\left|\sum_{i=1,2,3,4} |U_{ei}|^2\exp\left(i\frac{\Delta m^2_{i1}L}{2E_{\nu}}\right)\right|^2~, \\
&=& 1 - \sum_{i<j}4|U_{ei}|^2|U_{ej}|^2\sin^2\left(\frac{\Delta m^2_{ij}L}{4E_{\nu}}\right)~,
\end{eqnarray}
where $L$ is the antineutrino propagation distance (baseline), $E_{\nu}$ is the neutrino energy and $\Delta m^2_{ij}\equiv m_j^2-m_i^2$.

Current data constrain $|U_{e4}|^2<$~few~$\times 10^{-2}$ for $|\Delta m^2_{4i}|\gtrsim 10^{-2}$~eV$^2$  \cite{Cirelli:2004cz}, while $|\Delta m^2_{13}|\sim (2-3)\times 10^{-3}$~eV$^2$ and $\Delta m^2_{12}\sim 7.5\times 10^{-5}$~eV$^2$. We are interested in $L<2$~km and $E_{\nu}>2$~MeV so that 
\begin{equation}
\frac{\Delta m^2_{12} L}{4E_{\nu}}<0.1~.
\end{equation}
Finally, $|U_{e1}|^2\sim 0.7$, $|U_{e2}|^2\sim 0.3$, and $|U_{e3}|^2<0.04$. To a good approximation, for $L$ and $E_{\nu}$ values of interest,
\begin{equation}
P_{ee}=1-4(1-|U_{e3}|^2-|U_{e4}|^2)|U_{e3}|^2\sin^2\left(\frac{\Delta m^2_{13}L}{4E_{\nu}}\right)-4(1-|U_{e4}|^2)|U_{e4}|^2\sin^2\left(\frac{\Delta m^2_{14}L}{4E_{\nu}}\right)~,
\end{equation}
where we dropped a term proportional to
\begin{equation}
4|U_{e3}|^2|U_{e4}|^2\left[\sin^2\left(\frac{\Delta m^2_{34}L}{4E_{\nu}}\right)-\sin^2\left(\frac{\Delta m^2_{14}L}{4E_{\nu}}\right)\right]~,
\label{eq:43-41}
\end{equation}
which is suppressed by four powers of small mixing angles and vanishes in the limit $\Delta m^2_{13}L/4E_{\nu}\ll 1$, and a term proportional to 
\begin{equation}
4|U_{e1}|^2|U_{e2}|^2\sin^2\left(\frac{\Delta m^2_{12}L}{4E_{\nu}}\right)\simeq8\times 10^{-3}\left(\frac{L}{\rm 2~km}\right)^2\left(\frac{\rm 2~MeV}{E_{\nu}}\right)^2~.
\end{equation}
This contribution is both very small for almost all baselines and energies of interest and also well-known from combined solar and KamLAND data. Given the intentions of this paper, its inclusion in the following discussion is of no practical consequence.\footnote{To be very concrete, one can always subtract out these very small ``solar'' effects from the measured $P_{ee}$ and perform the analysis in the ``solar-subtracted'' data sample.} We have also not included matter effects, which can be safely neglected. 

 In this limit, $P_{ee}$ is a function of two mass-squared differences ($\Delta m^2_{14}$ and $\Delta m^2_{13}$) and two elements of the lepton mixing matrix. We will parameterize $|U_{e3}|$ and $|U_{e4}|$ in the ``standard way'' (see, for example, \cite{deGouvea:2008nm}):
\begin{equation}
|U_{e3}|^2=\cos^2\theta_{14}\sin^2\theta_{13}~,~~~~|U_{e4}|^2=\sin^2\theta_{14}~,
\end{equation}
so that 
\begin{equation}
P_{ee}=1-\cos^4\theta_{14}\sin^22\theta_{13}\sin^2\left(\frac{\Delta m^2_{13}L}{4E_{\nu}}\right)-\sin^22\theta_{14}\sin^2\left(\frac{\Delta m^2_{14}L}{4E_{\nu}}\right)~.
\end{equation}
In the limit $\theta_{14}\to 0$ we recover the well-known expression for $P_{ee}$ at $\theta_{13}$-driven reactor neutrino experiments: $P_{ee}=1-\sin^22\theta_{13}\sin^2(\Delta m^2_{13}L/4E_{\nu})$. Note that, in spite of the fact that we have three neutrino flavors, $P_{ee}$ is insensitive to the ``sterile mass hierarchy'', {\it i.e.}, it cannot tell whether $m_4>m_{3,2,1}$ or $m_4<m_{3,2,1}$. The reason is that we are not sensitive to the contribution Eq.~(\ref{eq:43-41}).

While we have information regarding $\Delta m^2_{13}$, nothing is known about $\Delta m^2_{14}$. Here we concentrate on the region $\Delta m^2_{14}\in [1-10]\times 10^{-2}$~eV$^2$, for different reasons. Phenomenologically, we are interested in sterile neutrino effects that qualitatively modify the observed oscillation pattern. $\Delta m^2_{14}$ values smaller than $10^{-2}$~eV$^2$ start to mimick $\Delta m^2_{13}$ effects and are not considered, while $\Delta m^2_{14}$ values much larger than $10^{-1}$~eV$^2$ lead to averaged out effects at both the near and far detectors and are hence less prominent. A detailed study of the effect of LSND/Mini-BooNE-inspired sterile neutrinos ($\Delta m^2_{14}\gtrsim 1$~eV$^2$) was recently presented in \cite{Bandyopadhyay:2007rj}. Theoretically, we have in mind a simple ``seesaw'' Lagrangian for the sterile neutrinos, where mostly active neutrino masses, mostly sterile neutrino masses and active--sterile mixing angles are naively related via $\theta_{14}^2\sim U^2(m_3/m_4)$, where $U$ stands for some weighted linear combination of the ``active'' mixing angles. Therefore $\theta_{14}^2$ values around $10^{-1}$ or $10^{-2}$ are naively related to $m_4$ values which are not more than one or two orders of magnitude larger than $m_{3,2,1}<0.05$~eV. Experimentally, the Bugey experiment disfavors $\sin^22\theta_{14}$ values smaller than $4\times 10^{-2}$ for $\Delta m^2_{14}\gtrsim 10^{-1}$~eV$^2$ while cosmological considerations constrain $\sin^2\theta_{14}<10^{-2}$ for $\Delta m^2_{14}>10^{-1}$~eV$^2$ \cite{Cirelli:2004cz}. Both constraints are significantly alleviated for smaller values of $\Delta m^2_{14}$. For $\Delta m^2_{14}=10^{-2}$~eV$^2$, $\sin^22\theta_{14}$ values as large as $10^{-1}$ are allowed by all neutrino and cosmological/astrophysical data \cite{Cirelli:2004cz}.

To qualitatively understand the effect of sterile neutrinos in reactor neutrino setups we look at
\begin{equation}
\frac{\Delta m^2_{14}L}{4E_{\nu}}=1.267 \left( \frac{\Delta m^2_{14}}{10^{-2}~\rm eV^2}\right)\left(\frac{L}{400~\rm m}\right)\left(\frac{4~\rm MeV}{E_{\nu}}\right).
\end{equation}
For reactor antineutrinos, the oscillation length associated to $\Delta m^2_{14}$ is of order the near detector distance for $\Delta m^2_{14}$ values in the range of interest. This means that sterile neutrinos can affect $P_{ee}$ in the near and far detectors in distinct ways. Given that the sensitivity to very small values of $\theta_{13}$ relies on a ``near versus far'' comparison, the presence of such sterile neutrinos can impact the reach of these experimental setups. Before proceeding with more quantitative results, we present a concrete qualitative example of what we mean.

Imagine that in order to rule out a particular $\theta_{13}$ value we relied solely on whether the near/far ratio deviated from expectations. It is useful to define the observable ${\rm near/far}\equiv (N_{\rm near}/N^0_{\rm near})/(N_{\rm far}/N^0_{\rm far})$, where $N$ is the observed number of events and $N^0$ the expected number of events (in the absence of oscillations) in the near or far detectors. 
\begin{equation}
\frac{\rm near}{\rm far}\sim\frac{\overline{P}_{ee}(\rm near)}{\overline{P}_{ee}(\rm far)}~,
\end{equation}  
where $\overline{P}_{ee}$ indicates the average electron antineutrino survival probability. In the absence of oscillations ${\rm near/far}=1$. Ignoring $\Delta m^2_{13}$ effects in  the near detector and assuming that in the far detector the average value of $\sin^2\Delta m^2_{14}L/4E_{\nu}=1/2$,
\begin{equation}
\frac{\rm near}{\rm far}\sim \frac{1-\sin^22\theta_{14}a_{14}(\rm near)}{1-\cos^4\theta_{14}\sin^22\theta_{13}a_{13}(\rm far)-0.5\sin^22\theta_{14}}~,
\end{equation}  
where $a_{ij}$ is the average values of $\sin^2(\Delta m^2_{ij}L/4E_{\nu})$ in the near or the far detector. A measurement of near/far consistent with one would be consistent with nonzero $\theta_{13}$ values satisfying 
\begin{eqnarray}
\sin^22\theta_{14}a_{14}(\rm near)&=&\cos^4\theta_{14}\sin^22\theta_{13}a_{13}(\rm far)+0.5\sin^22\theta_{14}~, \\
\sin^22\theta_{13}&=&\frac{\sin^22\theta_{14}}{\cos^4\theta_{14}}\left(\frac{a_{14}(\rm near)-0.5}{a_{13}(\rm far)}\right)~. 
\label{eq:estimate}
\end{eqnarray}
$\theta_{13}$ effects in the far detector can be ``compensated'' by $\theta_{14}$ effects in the near detector as long as $a_{14}(\rm near)$ is larger than one half and $\theta_{14}$ is large enough. If $a_{13}$(far)$\sim a_{14}$(near)$\sim 1$ such an effect occurs if $\sin^22\theta_{13}\sim \sin^22\theta_{14}/2$. In summary, the absence of a discrepant near/far ratio can be interpreted in one of two ways: either there are only three neutrinos  and $\theta_{13}$ is very small, or there are four neutrinos and $\Delta m^2_{14}$ and $\theta_{14}$ are such that $\theta_{13}$ effects in the near detector and $\theta_{14}$ effects in the far detector ``cancel!'' 

The results of the above simplified analysis need to be qualified. In the case of ``large'' $\Delta m^2_{14}$, $a_{14}(\rm near)\to 1/2$, and the ambiguity is erased (this is the scenario discussed in \cite{Bandyopadhyay:2007rj}). More importantly, in the case of large enough $\theta_{13}$ and $\theta_{14}$ (or, equivalently, once a large enough data sample is available), one expects to extract most of the information from distortions (or lack thereof) in the electron antineutrino energy spectrum. Ultimately, we expect that $\theta_{14}$ effects will lead to a slight loss of sensitivity to the smallest values of $\theta_{13}$, {\it i.e.}, those close to the $\theta_{13}$ sensitivity boundaries computed in \cite{Ardellier:2006mn,Guo:2007ug}. 

Nonzero values of $\theta_{14}$, on the other hand, may lead to potentially very unexpected results. For example, it is easy to see that near/far ratio may exceed one (not possible when $\theta_{14}=0$). In the case of $\theta_{13}=0$, near/far is proportional to 
\begin{equation}
\frac{\rm near}{\rm far}\sim\frac{1-\sin^22\theta_{14}a_{14}(\rm near)}{1-0.5\sin^22\theta_{14}}~.
\end{equation}
In this case, a near/far result different from one leads to a measurement of both $\sin^22\theta_{14}$ and $\Delta m^2_{14}$. 

\setcounter{equation}{0}
\section{Numerical Results}
\label{sec:numerical}

In order to study the impact of very light sterile neutrinos on $\theta_{13}$-driven reactor antineutrino experiments, we simulate data for different values of the mixing parameters $\Delta m^2_{13,14}$ and $\theta_{13,14}$ and proceed to analyze these data under distinct hypotheses. The expected number of events $N_i$ at a detector located a distance $L$ from the source (assumed to be point-like) which are associated to incoming neutrino energies between $E_i$ and $E_{i}+\Delta E$ is
\begin{equation}
N_i(L)=N_0(L)\int_{E_i}^{E_{i}+\Delta E}\phi(E_{\nu})\sigma(E_{\nu})P_{ee}(E_{\nu}){\rm d}E_{\nu},
\label{eq:ni}
\end{equation}
where $\phi(E_{\nu})$ is the energy dependent antineutrino flux while $\sigma(E_{\nu})$ is the total cross-section for $\bar{\nu}_e+p\to e^++n$. For concreteness, we use the expression for the time-averaged $\phi(E_{\nu})$ adopted  in \cite{Huber:2004xh} (see also \cite{Murayama:2000iq}) while we use the expression for $\sigma(E_{\nu})$ computed in \cite{Vogel:1999zy}. $N_0(L)$ is a normalization factor that depends on the size of the detector, the intensity of the source, the running time and the source--detector distance ($N_0(L)\propto L^{-2}$).

We analyze our different simulated data sets by performing a simple $\chi^2$ test, where
\begin{equation}
\chi^2(\Delta m^2_{14},\theta_{14},\Delta m^2_{13},\theta_{13},\alpha)=\left.\sum_{i=1}^{n_{\rm bins}}\frac{(N_i-(1+\alpha)T_i)^2}{(\delta N_i)^2}\right|_{\rm near}+\left.\sum_{i=1}^{n_{\rm bins}}\frac{(N_i-(1+\alpha)T_i)^2}{(\delta N_i)^2}\right|_{\rm far}+\frac{\alpha^2}{(\delta \alpha)^2}~.
\label{eq:chisq}
\end{equation}
Here $T_i$ is the theoretically expected number of events in the $i$-th energy bin, given by Eq.~(\ref{eq:ni}). $T_i$ depends on the oscillation parameters $\Delta m^2_{14},\theta_{14},\Delta m^2_{13},\theta_{13}$. $\delta N_i$ is the error on $N_i$, which we assume is purely statistical and Gaussian: $\delta N_i=\sqrt{N_i}$. The sums are performed over all $n_{\rm bins}$ energy bins in the near and far detectors. $\alpha$ is a nuisance parameter that governs how well the expected number of reactor antineutrino induced events at zero baseline can be predicted. $\delta \alpha$ contains all uncertainties that are common to the expected number of events at both the near and far detectors, including uncertainties on the overall reactor antineutrino flux, uncertainties on the energy dependence of the flux, uncertainties on the antineutrino--target cross-sections, etc. No other systematic effects are considered, and, for the purposes of this analysis, we assume that $\alpha$ does not depend on the energy bin. The role of the $\alpha$ parameter is simple. Roughly speaking, when  $T_i/N_i-1$ is smaller than $\delta\alpha$, most of the statistical power in the analysis comes from a comparison of near detector versus far detector expectations since in the limit where either detector is ``turned off'' one can choose $\alpha$ such that $(N_i-(1+\alpha)T_i)^2$ is small while $\alpha^2/(\delta\alpha)^2$ is order one. 

We consider two different simulated setups. In the Double-CHOOZ-like setup \cite{Ardellier:2006mn}, we choose $L_{\rm near}=400$~m, $L_{\rm far}=1050$~m and $N_0(L)$ such that one would accumulate 40,000 events in the far detector in the absence of oscillations. For the near detector, we set the number of unoscillated events equal to that in the far detector times $L^2_{\rm far}/L^2_{\rm near}$.\footnote{In the Double CHOOZ experiment, the near detector is expected to be smaller than the far detector. However, since most of the results presented here are dominated by the statistics in the far detector, we find that this discrepancy does not lead to any qualitatively distinct results. The same comment applies to the Daya-Bay-like setup.}  In the Daya-Bay-like setup \cite{Guo:2007ug}, we choose $L_{\rm near}=400$~m, $L_{\rm far}=1800$~m and $N_0(L)$ such that one would accumulate 75,000 events in the far detector in the absence of oscillations. As in the Double-CHOOZ-like setup, we set the unoscillated number of events in the near detector equal to that in the far detector times $L^2_{\rm far}/L^2_{\rm near}$. We fix $\delta\alpha=3\%$, which is of order the overall systematic uncertainty estimated in the CHOOZ experiment \cite{Apollonio:2002gd}. Both the Double-CHOOZ and Daya Bay collaborations are aiming at understanding their experimental setups at a level that translates into, roughly, $\delta\alpha\sim 1\%$. In what follows, we briefly comment on the impact of allowing for smaller $\delta\alpha$ values.

\begin{figure}[ht]
\centerline{\includegraphics[scale=0.65]{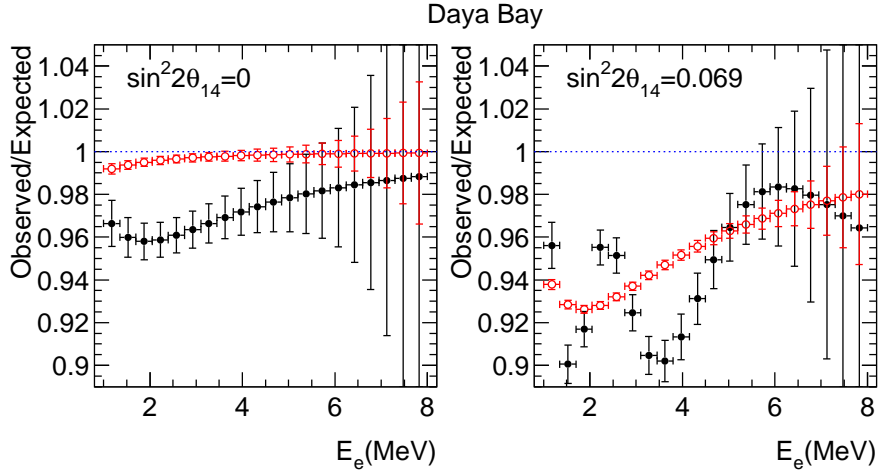}}
\caption{Number of events per energy bin in the Daya-Bay-like setup, normalized to the expected number of events in the absence of oscillations. Error bars are statistical only. The ``data'' correspond to $\sin^22\theta_{13}=0.042$, $\Delta m^2_{13}=2.2\times 10^{-3}$~eV$^2$, $\Delta m^2_{14}=1.0\times 10^{-2}$~eV$^2$ and $\sin^22\theta_{14}=0$ (left-hand side) or $\sin^22\theta_{14}=0.069$ (right-hand side). The grey [red] open circles (black closed circles) with smaller (larger) error bars correspond to ``data'' in the near (far) detector. The dotted [blue] line indicates the no-oscillation case.}
\label{fig:data}
\end{figure}

Two sample ``data sets'' for the Daya-Bay-like setup can be found in Fig.~\ref{fig:data}. It depicts the number of events normalized to the expected number of events in the absence of oscillations in twenty equal-width recoil positron kinetic energy bins ($E_e=E_{\nu}-1.293$~MeV) between 1 and 8~MeV. In both panels, $\sin^22\theta_{13}=0.042$, $\Delta m^2_{13}=2.2\times 10^{-3}$~eV$^2$, $\Delta m^2_{14}=1.0\times 10^{-2}$~eV$^2$. In the left-hand side, $\theta_{14}=0$ so that, as far as this observable ($P_{ee}$ at the baselines and energies of interest here) is concerned, there are no sterile neutrinos. In the right-hand side, $\sin^22\theta_{14}=0.069$. In the case $\sin^22\theta_{14}=0.069$ two features are noteworthy. One is that the $\Delta m^2_{14}$ effects lead to visible distortions in the recoil electron energy spectrum both in the neat and far detectors, assuming the energy resolution is such that one can ``see'' the binning depicted in the figure. The other is that, on average, the near and far detectors point to a similar suppression of the expected number of events, as discussed in Sec.~\ref{sec:mixing}.

\subsection{No Evidence for Oscillations}

In the absence of an oscillation signal, next-generation $\theta_{13}$-driven experiments rule out regions of the $\sin^22\theta_{13}\times\Delta m^2_{13}$ plane that are currently allowed by all neutrino data. Such a result would severely impact planning for next and next-next generation neutrino experiments. It would, for example, reveal that the NO$\nu$A experiment \cite{Ayres:2004js} cannot determine the neutrino mass hierarchy and strengthen the case for building a muon storage ring (neutrino factory) \cite{Bandyopadhyay:2007kx}. 

In order to study the impact of sterile neutrinos, we simulate data (as described above) consistent with no oscillations ($\theta_{13}=\theta_{14}=0$) and analyze it under two distinct hypotheses: (i) there are no light sterile neutrinos (as far as the setups in question are concerned, this is equivalent to $\theta_{14}=0$) and (ii) there is a fourth neutrino mass state with $\Delta m^2_{14}\in[1,10]\times 10^{-2}$~eV$^2$ and  $\sin^22\theta_{14}<0.1$. In either case, we restrict $\Delta m^2_{13}\in[2,3]\times 10^{-3}$~eV$^2$, as dictated by current neutrino data.\footnote{Our $\Delta m^2_{13}$ window agrees with the 90\% confidence level allowed range quoted in the particle data book \cite{Amsler:2008zz} and is much wider than the most recent MINOS result, $\Delta m^2_{13}\in[2.17,2.69]\times 10^{-3}$~eV$^2$ at the $2\sigma$ level \cite{Adamson:2008zt}.}

\begin{figure}[ht]
\centerline{\includegraphics[scale=0.65]{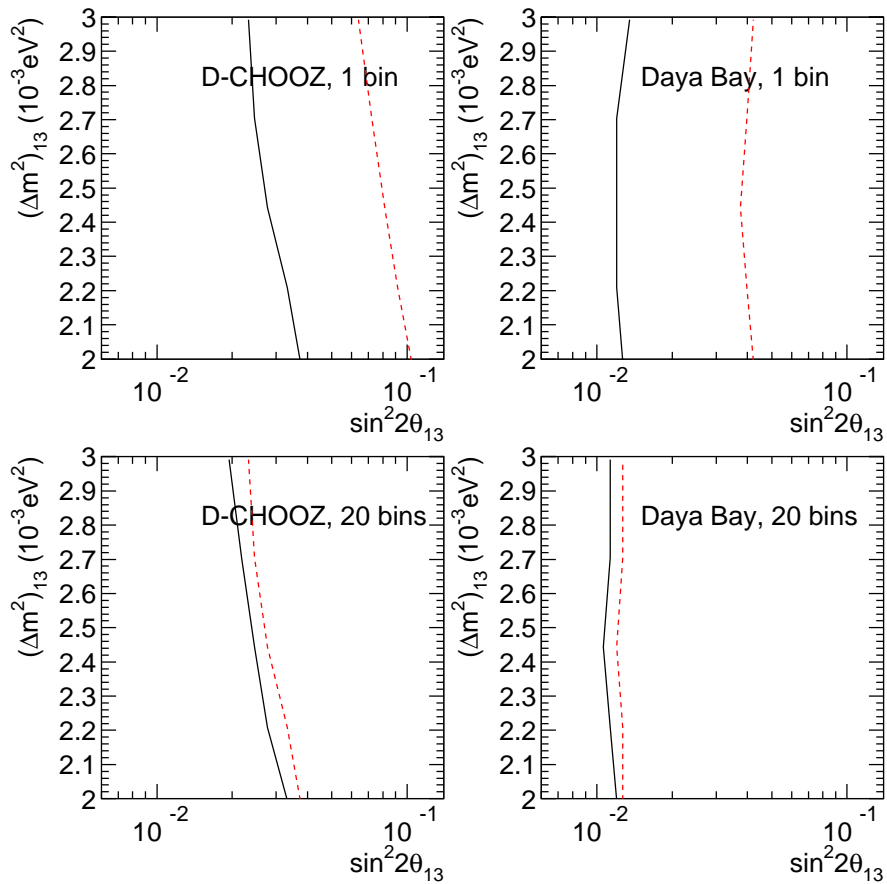}}
\caption{Region of the $\sin^22\theta_{13}\times\Delta m^2_{13}$ parameter ruled out, at the $2\sigma$ confidence level in the Double-CHOOZ-like setup (left-hand side) and the Daya-Bay like setup (right-hand side). In the top panels we depict the result of a ``total rate'' experiment ($n_{\rm bins}=1$) while in the bottom the ``data'' is subdivided into 20 recoil positron energy bins ($n_{\rm bins}=20$). The region to the right of the black, continuous line is ruled out under the hypothesis that there are no sterile neutrinos ($\theta_{14}=0$). The region to the right of the gray (red) dashed curve is ruled out under the hypothesis that $\Delta m^2_{14}\in[1,10]\times 10^{-2}$~eV$^2$ and $\sin^22\theta_{14}<0.1$.}
\label{fig:sens_13}
\end{figure}
Figure~\ref{fig:sens_13} depicts the region of parameter space ruled out at the $2\sigma$ confidence level in the Double-CHOOZ-like (left-hand side) and Daya-Bay-like (right-hand side) setup. The darker continuous boundaries are obtained under the hypothesis that $\theta_{14}=0$. Note that, in spite of the simplified nature of our analyses, our results agree qualitatively with those in   \cite{Ardellier:2006mn,Guo:2007ug}. The lighter [red] dashed boundaries are obtained once $\chi^2$ is marginalized over the ``allowed'' $\sin^22\theta_{14}\times \Delta m^2_{14}$ parameter space. 

Throughout, we perform two different ``types'' of data analysis. In one case we consider a simple counting experiment ($n_{\rm bins}=1$),  {\it i.e.}, one counts how many electron antineutrino candidate events appear in the near and far detectors and compares these numbers against expectations. In this case,  the ability of sterile neutrinos to ``mask'' $\theta_{13}$ effects is optimal. The reason is simple. For a given value of $\sin^22\theta_{13},\Delta m^2_{13}$ one can  ``always'' find a value of $\sin^22\theta_{14},\Delta m^2_{14}$ in the allowed region such that the number of events in the near detector differs from expectations as much as the number of events in the far detector. In this case, one cannot rely on the near/far comparison to ``measure'' the expected number of neutrino events and the sensitivity is governed by external uncertainties ($\delta\alpha$ parameter in Eq.~(\ref{eq:chisq})). 

The other case under consideration is $n_{\rm bins}=20$, {\it i.e.}, we take into account not only the overall suppression of the electron antineutrino flux but also potential distortions of the recoil positron energy spectrum. 20 recoil positron energy bins between 1 and 8 MeV corresponds to $\Delta E= 350$~keV. This is slightly wider than the expected energy resolution at both Double-CHOOZ \cite{Ardellier:2006mn} and Daya Bay \cite{Guo:2007ug}. In this case, Fig.~\ref{fig:sens_13} reveals that the ability to exclude $\sin^22\theta_{13}$ values is not severely compromised by the light sterile neutrino hypothesis. The $n_{\rm bins}=20$ case is, perhaps, a more faithful estimate of the results one would obtain with a realistic detector simulation. Detector and energy dependent systematic effects (not included in our analyses), however, tend to reduce the power of the binned analysis compared to overall flux one.

\begin{figure}[ht]
\centerline{\includegraphics[scale=0.65]{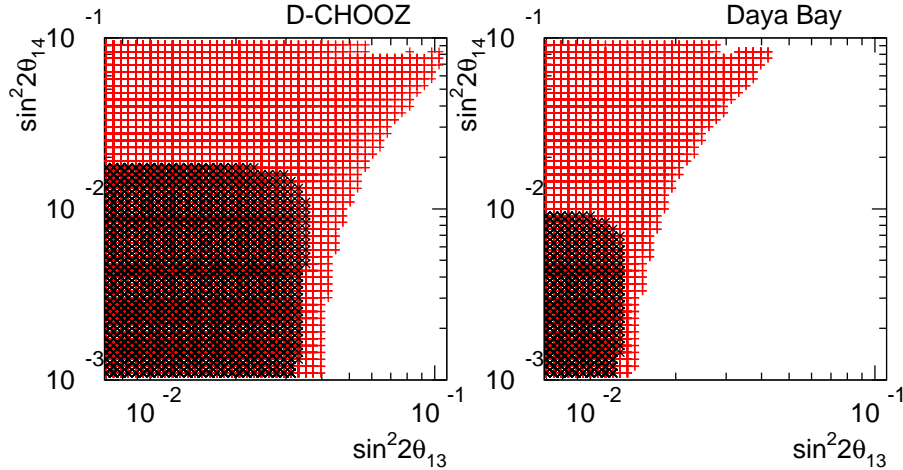}}
\caption{Allowed region of parameter space ($2\sigma$ confidence level) in the $\sin^22\theta_{13}\times \sin^22\theta_{14}$ plane assuming the data in the Double-CHOOZ-like setup (left-hand side) and in the Daya-Bay-like setup (right-hand side) are consistent with no oscillations. The (red) horizontal-vertical hatching indicates the result of a ``total rate'' experiment ($n_{\rm bins}=1$) while the black criss-crossed hatching corresponds to the  ``data'' subdivided into 20 recoil positron energy bins ($n_{\rm bins}=20$).}
\label{fig:sens_13_14}
\end{figure}

Figure~\ref{fig:sens_13_14} depicts the allowed region of parameter space in the $\sin^22\theta_{13}\times \sin^22\theta_{14}$ plane assuming the data in the Double-CHOOZ-like setup (left-hand side) and in the Daya-Bay-like setup (right-hand side) are consistent with no oscillations and after $\chi^2$ is marginalized over the allowed values of $\Delta m^2_{13}$ and $\Delta m^2_{14}$. In the counting experiment case (horizontal-vertical grey [red] hatching) one sees that when larger $\sin^22\theta_{14}$ values are considered, larger $\theta_{13}$ values are consistent with no oscillations. This is in agreement with the estimate made in Eq.~(\ref{eq:estimate}), which can be roughly translated as follows. By allowing different values of $\Delta m^2_{14}$, the near/far ratio, on average, can be set to one for all $\theta_{14}$ values satisfying $\sin^22\theta_{14}\le 2\sin^22\theta_{13}$, assuming $a_{13}(\rm far)\sim 1$. Note that if $\theta_{13}$ were set to zero, the analysis of a counting experiment consistent with no oscillations cannot rule out $\sin^22\theta_{14}=0.1$. This is due to the fact that for $n_{\rm bins}=1$ and large enough $\Delta m^2_{14}$ $\theta_{14}$-driven oscillations average out at both the near and the far detectors. In this case the sensitivity to $\sin^22\theta_{14}$ is dominated by $\delta \alpha$. Since that was set to $3\%$, in qualitative agreement with the original CHOOZ experiment, the sensitivity to $\sin^22\theta$ is similar to the sensitivity of the CHOOZ experiment to averaged out oscillations, $\sin^22\theta\lesssim 0.1$. For smaller values of $\delta \alpha$, values of $\sin^22\theta_{14}$ larger than several percent are ruled out even in the $n_{\rm bins}=1$ case. 

In the case of the more finely binned data (20 bins, criss-crossed black hatching) the absence of distortions in the energy spectrum at both detectors prevent large $\theta_{13}$ or $\theta_{14}$ values. Since $\theta_{14}$-driven oscillations lead, in the near detector,  to potentially visible distortions of the antineutrino energy spectrum even for the largest considered value of $\Delta m^2_{14}$, the upper bound on $\sin^22\theta_{14}$ is stronger than that on $\sin^22\theta_{13}$. For smaller values of $\delta \alpha$ the results associated to $n_{\rm bins}=20$ do not change qualitatively, while those associated to $n_{\rm bins}=1$ start to approach the $n_{\rm bins}=20$ case when $\delta\alpha\sim 1\%$.

\subsection{Evidence for Oscillations ( $\theta_{13}$-Driven)}

If $\theta_{13}$ is large ($\sin^22\theta_{13}=$~few$\times10^{-2}$), one expects a statistically significant disappearance of electron antineutrinos in the far detector of $\theta_{13}$-driven reactor antineutrino experiments. In this case, one expects to not only reject the $\theta_{13}=0$ hypothesis but also measure $\theta_{13}$ (and, to a much lesser extend, $\Delta m^2_{13}$). The result of such a measurement is also affected by whether one hypothesizes the presence of light sterile neutrinos. 

\begin{figure}[ht]
\centerline{\includegraphics[scale=0.65]{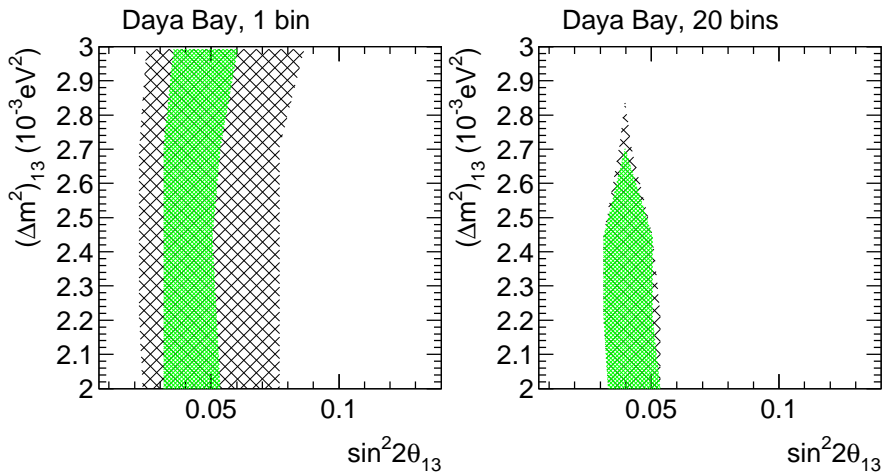}}
\caption{Allowed region of parameter space ($2\sigma$ confidence level) in the $\Delta m^2_{13}\times \sin^22\theta_{13}$ plane assuming the data in the Daya-Bay-like setup are consistent with $\Delta m^2_{13}=2.2\times 10^{-3}$~eV$^2$, $\sin^22\theta_{13}=0.042$ and $\theta_{14}=0$. The light [green] solid densely shaded region corresponds to analyzing the data assuming $\theta_{14}\equiv 0$, while the dark hatched region is obtaned if one allows for a fourth light neutrino. On the left panel a ``total rate'' experiment is performed ($n_{\rm bins}=1$) while in the right panel  the data was subdivided into 20 recoil positron energy bins ($n_{\rm bins}=20$).}
\label{fig:mea_13}
\end{figure}

As in the previous subsection, we consider both the $n_{\rm bins}=1$ and $n_{\rm bins}=20$ case in order to highlight the effect of the light  sterile neutrino hypothesis. Fig.~\ref{fig:mea_13} depicts the allowed region of parameter space extracted in the Daya-Bay-like setup assuming $\Delta m^2_{13}=2.2\times 10^{-3}$~eV$^2$, $\sin^22\theta_{13}=0.042$ and $\theta_{14}=0$ (the ``data'' depicted in Fig.~\ref{fig:data}(left)). The case $n_{\rm bins}=1$ ($n_{\rm bins}=20$) is depicted in the left (right) panel. Very similar results apply for the Double-CHOOZ-like setup.\footnote{Henceforth, we restrict our discussions to the Daya-Bay-like setup, keeping in mind that parallel results for the Double-CHOOZ-like setup are qualitatively very similar.}     

\begin{figure}[ht]
\centerline{\includegraphics[scale=0.65]{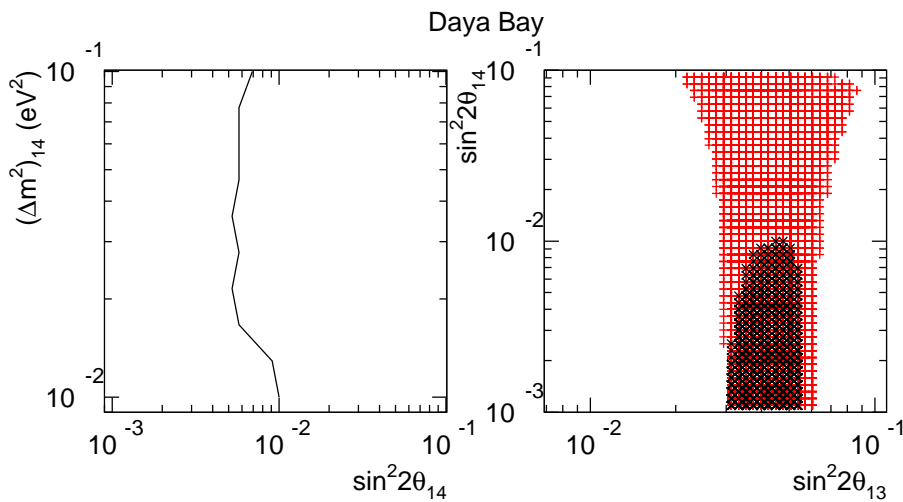}}
\caption{Allowed region of parameter space in the $\Delta m^2_{14}\times \sin^22\theta_{14}$ plane (left) and the $\sin^22\theta_{14}\times \sin^22\theta_{13}$ plane (right) assuming the data in the Daya-Bay-like setup are consistent with $\Delta m^2_{13}=2.2\times 10^{-3}$~eV$^2$, $\sin^22\theta_{13}=0.042$ and $\theta_{14}=0$ at the $2\sigma$ confidence level. On the right panel the (red) horizontal-vertical hatching indicates the result of a ``total rate'' experiment ($n_{\rm bins}=1$) while the black criss-crossed hatching corresponds to the  ``data'' subdivided into 20 recoil positron energy bins ($n_{\rm bins}=20$). On the left panel $n_{\rm bins}=20$.}
\label{fig:14_limit}
\end{figure}

Figure~\ref{fig:mea_13} reveals that, in the case of a simple counting experiment, whether or not one allows for a light sterile state significantly affects the precision with which $\theta_{13}$ can be measured. In the $n_{\rm bins}=20$ case, on the other hand, allowing for the existence of a light sterile neutrino does not significantly impact the precision with which $\theta_{13}$ (and $\Delta m^2_{13}$) is measured. The reason for this is simple. In the case of 1 bin, a larger or smaller value of $\sin^22\theta_{13}$ can be made consistent with the data if it is accompanied by a large enough hypothetical $\sin^2 2\theta_{14}$ value.  In the case $n_{\rm bins}=20$, large values of $\sin^22\theta_{14}$ are ruled out by the lack of distortion in the near (and far) detector recoil positron energy spectrum. This phenomenon is clearly seen in Fig.~\ref{fig:14_limit}(right), which depicts the allowed region of the $\sin^22\theta_{14}\times \sin^22\theta_{13}$ parameter space (once one marginalizes over the allowed $\Delta m^2_{13}$ and $\Delta m^2_{14}$ values). It is interesting to note that in the $n_{\rm bins}=1$ case  $\sin^22\theta_{14}$ values as large as 0.1 are allowed at the two-sigma confidence level. The reason for this is that, for large enough $\Delta m^2_{14}$ and $n_{\rm bins}=1$, the sensitivity to $\sin^22\theta_{14}$ is dominated by $\delta \alpha$, as discussed in the previous subsection. For smaller values of $\delta\alpha$ we find that ``large'' values of $\sin^22\theta_{14}$ are ruled out even in the $n_{\rm bins}=1$ case.

In the case $n_{\rm bins}=20$, the absence of a distorted positron energy spectrum in the near detector rules out $\sin^22\theta_{14}$ values larger than $10^{-2}$ or so. Fig.~\ref{fig:14_limit}(left)  depicts the region of the  $\Delta m^2_{14}\times \sin^22\theta_{14}$ plane ruled out at the two sigma level assuming the data in the Daya-Bay-like setup are consistent with $\Delta m^2_{13}=2.2\times 10^{-3}$~eV$^2$, $\sin^22\theta_{13}=0.042$ and $\theta_{14}=0$. Such a result would improve the current constraints on $\theta_{14}$ by about an order of magnitude for the values of $\Delta m^2_{14}$ highlighted here. 

\subsection{Evidence for Oscillations ($\theta_{13}$ and $\theta_{14}$-Driven)}

If $\theta_{14}$ is nonzero (and large enough) and $\Delta m^2_{14}\sim~{\rm few}\times 10^{-2}$~eV$^2$, $\theta_{14}$-effects will produce non-trivial, distinct effects in the near and far detectors, as depicted in Fig.~\ref{fig:data}(right). In this case, allowing for the presence of a light sterile neutrino during the data analysis would prove to be more than a choice --- it would be necessary in order to obtain a proper fit to the data of the reactor neutrino experiments under investigation here.  

\begin{figure}[ht]
\centerline{\includegraphics[scale=0.65]{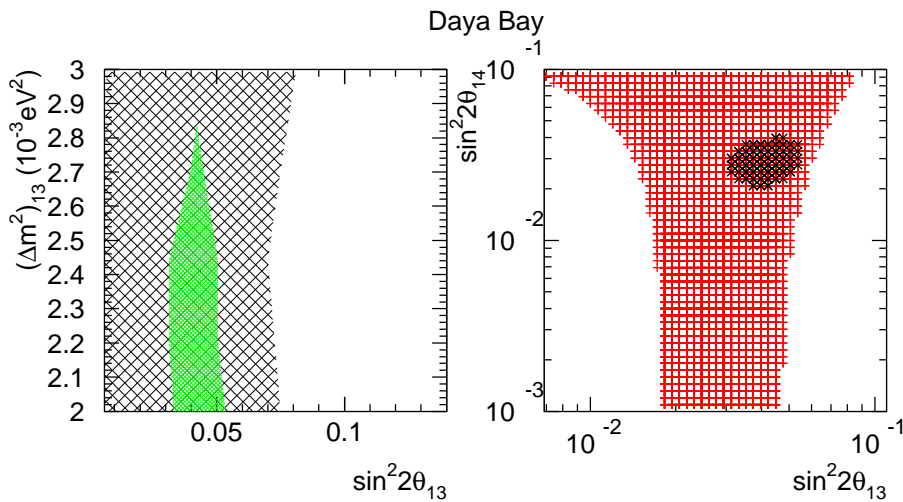}}
\caption{Allowed region of parameter space in the $\Delta m^2_{13}\times \sin^22\theta_{13}$ plane (left) and the $\sin^22\theta_{14}\times \sin^22\theta_{13}$ plane (right) assuming the data in the Daya-Bay-like setup are consistent with $\Delta m^2_{13}=2.2\times 10^{-3}$~eV$^2$, $\sin^22\theta_{13}=0.042$, $\Delta m^2_{14}=0.01$~eV$^2$, and $\sin^22\theta_{14}=0.03$ at the $2\sigma$ confidence level. On the left panel, the light [green] solid densely shaded region corresponds to $n_{\rm bins}=20$, while the dark hatched region corresponds to $n_{\rm bins}=1$. On the right panel, the (red) horizontal-vertical hatching indicates the result of a ``total rate'' experiment ($n_{\rm bins}=1$) while the black criss-crossed hatching corresponds to the  ``data'' subdivided into 20 recoil positron energy bins ($n_{\rm bins}=20$).}
\label{fig:14_13_mea}
\end{figure}

Figure~\ref{fig:14_13_mea} depicts two two-parameter 2$\sigma$ allowed regions of parameter space if the data in the Daya-Bay-like setup were consistent with $\Delta m^2_{13}=2.2\times 10^{-3}$~eV$^2$, $\sin^22\theta_{13}=0.042$, $\Delta m^2_{14}=0.01$~eV$^2$, and $\sin^22\theta_{14}=0.03$. Figure~\ref{fig:14_13_mea}(left) depicts how well $\sin^22\theta_{13}$ (and $\Delta m^2_{13}$) can be measured in the case $n_{\rm bins}=1$ and $n_{\rm bins}=20$ after one marginalizes over $\sin^22\theta_{14}$ and $\Delta m^2_{14}$. These should be compared to Fig.~\ref{fig:mea_13}. It is clear that in the $n_{\rm bins}=1$ case the fact that $\sin^22\theta_{14}$ is nonzero renders a ``rates-only'' measurement of $\sin^22\theta_{13}$ harder. In the $n_{\rm bins}=20$ case, on the other hand, the impact of a non-zero $\sin^22\theta_{14}$ when it comes to measuring $\sin^22\theta_{13}$ is small. The reason for this is that in the $n_{\rm bins}=20$ case distortions in the near (and, to a lesser extent, in the far) detector determine $\Delta m^2_{14}$ and $\sin^22\theta_{14}$,  allowing the near/far comparison to ``fully contribute'' to the measurement of $\theta_{13}$. 

Figure~\ref{fig:14_13_mea}(right) depicts how well $\sin^22\theta_{13}$ and $\sin^22\theta_{14}$ can be measured in the case $n_{\rm bins}=1$ and $n_{\rm bins}=20$ after one marginalizes over both mass-squared differences. In the case $n_{\rm bins}=1$, virtually no constraint can be set on $\sin^22\theta_{13}$. Curiously enough, if sterile effects were not included ($\theta_{14}\equiv0$), one would be able to establish that $\sin^22\theta_{13}\neq 0$. In this case, however, the wrong hypothesis in the data analysis would point to an allowed range for $\sin^22\theta_{13}$ that is slightly less than its real value (this is true at around the 1$\sigma$ level). For larger ``true'' values of $\theta_{14}$ this effect is more pronounced.  As discussed earlier, in the case $n_{\rm bins}=20$ one is able to obtain a precise measurement of both mixing angles. 

\begin{figure}[ht]
\centerline{\includegraphics[scale=0.65]{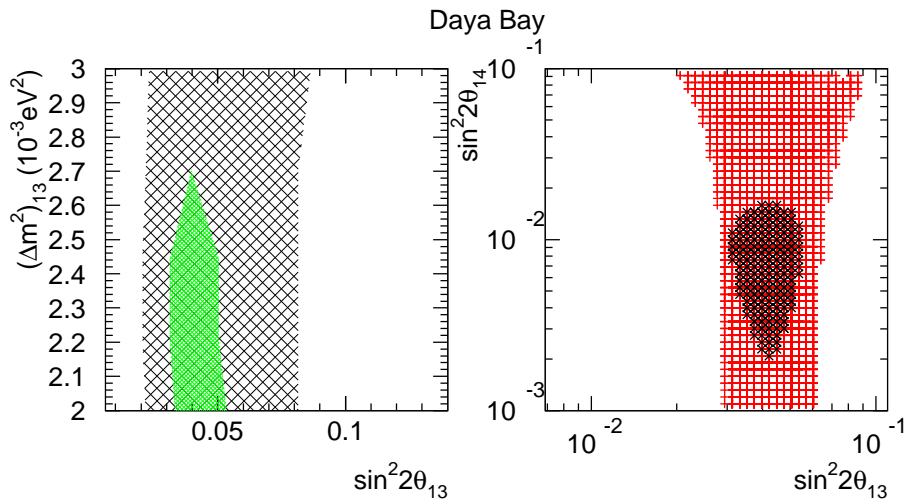}}
\caption{Allowed region of parameter space in the $\Delta m^2_{13}\times \sin^22\theta_{13}$ plane (left) and the $\sin^22\theta_{14}\times \sin^22\theta_{13}$ plane (right) assuming the data in the Daya-Bay-like setup are consistent with $\Delta m^2_{13}=2.2\times 10^{-3}$~eV$^2$, $\sin^22\theta_{13}=0.042$, $\Delta m^2_{14}=0.1$~eV$^2$, and $\sin^22\theta_{14}=0.0091$ at the $2\sigma$ confidence level. On the left panel, the light [green] solid densely shaded region corresponds to $n_{\rm bins}=20$, while the dark hatched region corresponds to $n_{\rm bins}=1$. On the right panel, the (red) horizontal-vertical hatching indicates the result of a ``total rate'' experiment ($n_{\rm bins}=1$) while the black criss-crossed hatching corresponds to the  ``data'' subdivided into 20 recoil positron energy bins ($n_{\rm bins}=20$).}
\label{fig:14_13_mea_2}
\end{figure}

For smaller values of true $\sin^22\theta_{14}$, the impact of sterile neutrinos is, of course, less pronounced. Figure~\ref{fig:14_13_mea_2} depicts two two-parameter 2$\sigma$ allowed regions of parameter space if the data in the Daya-Bay-like setup were consistent with $\Delta m^2_{13}=2.2\times 10^{-3}$~eV$^2$, $\sin^22\theta_{13}=0.042$, $\Delta m^2_{14}=0.1$~eV$^2$, and $\sin^22\theta_{14}=0.0091$. In this case, even in the $n_{\rm bins}=1$ case one can establish that $\sin^22\theta_{13}\neq 0$ (at the 2$\sigma$ level). In the $n_{\rm bins}=20$ case, a reasonable measurement of both $\sin^22\theta_{13}$ and $\sin^22\theta_{14}$ can be performed --- at the $2\sigma$ level, $\sin^22\theta_{14}\in[0.002,0.015]$ and $\sin^22\theta_{13}\in[0.03,0.055]$, and correlations are small.  

We conclude this section with a short comment on the case $\theta_{13}=0$ and $\theta_{14}\neq 0$ large. As far as measuring $\sin^22\theta_{14}$ is concerned, the situation here is qualitatively similar to the previous two cases displayed above. Due to the cancellation effects discussed earlier, an $n_{\rm bins}=1$ analysis fails to severely constrain either $\theta_{13}$ (similar to the situation depicted in Fig.~\ref{fig:sens_13}(top)) and $\theta_{14}$. An $n_{\rm bins}=20$ analysis, on the other hand, allows one to not only measure $\sin^22\theta_{14}$ and $\Delta m^2_{14}$ with good precision but also constrain $\sin^22\theta_{13}$ significantly, as in Fig.~\ref{fig:sens_13}(bottom).

\section{Discussion and Conclusions}
\label{sec:conclusion}

We have discussed the impact of very light sterile neutrinos on the interpretation of next-generation $\theta_{13}$-driven reactor antineutrino experiments. For $\Delta m^2_{14}$ values between 1 and 10 $\times 10^{-2}$~eV$^2$, $\theta_{14}$-driven oscillations affect data in the far and near detectors differently, leading to several potentially interesting effects. In the absence of a positive oscillation signal, the hypothesis $\theta_{14}\neq 0$ negatively impacts one's ability to rule out very small $\theta_{13}$ values as would-be $\theta_{14}$-driven effects in the near and far detectors can mask $\theta_{13}$-driven effects in the far detector. Similarly, if a positive $\theta_{13}$-driven signal is observed, would-be $\theta_{14}$-driven effects negatively impact ones ability to measure the value of $\sin^22\theta_{13}$. The above results are more or less pronounced depending on one's ability to ``see'' distortions of the recoil positron energy spectrum. A simple counting experiment is very susceptible to hidden $\theta_{14}$-driven effects while a binned data analysis (at both the near and far detectors) seems able to disentangle $\theta_{13}$ from $\theta_{14}$-driven effects. 

If there is indeed a light sterile neutrino associated to  $\Delta m^2_{14}\in [1,10]\times 10^{-2}$~eV$^2$, $\theta_{13}$-driven reactor antineutrino experiments should be able to see large spectral distortions in the near detector, in which case one can determine not only $\theta_{14}$ but also $\Delta m^2_{14}$. It is curious to note that, for $\Delta m^2_{14}$ values close to 0.1~eV$^2$, the roles of the near and far detectors as far as studying $\theta_{14}$-driven effects are reversed compared to those associated to studying $\theta_{13}$-driven effects. For these mass-squared differences, oscillation effects average out in the far detector (even in a finely-binned analysis). This allows one to determine the ``averaged-out'' neutrino flux and hence helps extract the value of $\theta_{14}$ from the depths of the oscillation minima. In the absence of $\theta_{14}$-driven effects, we estimate that values of $\sin^22\theta_{14}$ as small as 1 percent can be ruled out with a binned analysis (Fig.~\ref{fig:14_limit}(left)).
 
Sterile neutrino effects at $\theta_{13}$-driven reactor experiments have been considered in the past, but in a different mass-squared difference regime. The authors of  \cite{Bandyopadhyay:2007rj} recently discussed the effect of sterile neutrinos related to a potential $3+2$ solution to the LSND anomaly.\footnote{After MiniBooNE data became available, it was pointed out that a 3+2 solution to all short baseline appearance data exists. This solution, however, is disfavored by disappearance searches sensitive to the same oscillation frequencies \cite{Maltoni:2007zf}.} There, the new mixing angles and mass-squared differences lead to averaged out effects at both the near and far detectors. Here, on the other hand, we concentrate on a lower mass-squared difference regime where this is not the case.
 
If there are indeed such light sterile neutrinos, it is likely that the most sensitive terrestrial probes of their existence are the setups discussed here. This is not a coincidence: we are concentrating on values of $\Delta m^2_{14}$ where oscillation effects in the near detector are optimal. Other near-future probes include long-baseline accelerator-based neutrino oscillation experiments. However, $\Delta m^2_{14}$-driven effects at next-generation $\theta_{13}$-driven appearance long-baseline experiments are proportional to the product of the squares of two potentially small mixing angles ($P_{\mu e}\propto \theta_{14}^2\theta_{24}^2$, using the notation of \cite{deGouvea:2008nm}). Furthermore, next-generation studies of muon neutrino disappearance at similar setups are also unlikely to achieve sensitivity competitive with the one discussed here. Note that $\Delta m^2_{14}$ is large enough that it leads to averaged-out effects at the far detector but too small to mediate observable effects in the near detectors of next-generation long-baseline experiments.\footnote{In long-baseline experiments, $L_{\rm near}/L_{\rm far}\sim 10^{-3}$.} Detailed recent analyses of next-generation oscillation probes of light sterile neutrinos, concentrating on $\Delta m^2_{14}$ values above 0.1~eV$^2$, can be found in \cite{Donini:2007yf,Dighe:2007uf,Goswami:2008mi,Choubey:2007ji}. 
 
It is important to appreciate that this range of sterile neutrino parameters is not motivated by any existing data, nor does it help address any existing outstanding issue in fundamental physics. On the other hand, as already emphasized, little is known about sterile neutrinos. Evidence for sterile neutrinos at any mass range would qualitatively change our understanding of particle physics (and probably shed light on the mechanism behind light neutrino masses).  
 
Finally, light sterile neutrinos qualify as an example of non-standard physics that may appear in next-generation reactor neutrino experiments. They can not only lead to nontrivial oscillation patterns in the data, but also modify the interpretation of reactor antineutrino data when it comes to measuring or constraining $\theta_{13}$. Our results also highlight the fact that new and interesting results may come out of the data in the near detector \cite{Conrad:2004gw}, which may provide more information than the measurement of the ``$L=0$'' reactor antineutrino flux. We conclude by re-emphasizing that our simulations and analyses are not aimed at realistically describing experimental setups or quantitatively gauging their reach. Qualitatively, however, our results capture the nontrivial impact of light sterile neutrinos at $\theta_{13}$-driven reactor antineutrino experiments. We hope that our findings will prompt the collaborations to pursue quantitative estimates of the impact of the very light sterile neutrinos introduced here. 

\section*{Acknowledgments}

A preliminary version of the results discussed here was presented as TW's senior undergraduate thesis at Northwestern University.
We are happy to thank Heidi Schellman for feedback on TW's undergraduate senior thesis and for encouragement. This work is sponsored in part by DOE grant \# DE-FG02-91ER40684.


\begin{thebibliography}{99}

\bibitem{Ardellier:2006mn}
  F.~Ardellier {\it et al.}  [Double Chooz Collaboration],
  arXiv:hep-ex/0606025.

\bibitem{Guo:2007ug}
  X.~Guo {\it et al.}  [Daya Bay Collaboration],
  arXiv:hep-ex/0701029.

 \bibitem{TASI} For a pedagogical discussion and many references see, for example,  A.~de Gouv\^ea,
arXiv:hep-ph/0411274.
For recent comprehensive reviews see, for example, M.~C.~Gonzalez-Garcia and M.~Maltoni,
  Phys.\ Rept.\  {\bf 460}, 1 (2008);
A.~Strumia and F.~Vissani,
  arXiv:hep-ph/0606054.
See also \cite{Amsler:2008zz}.

\bibitem{seesaw} P.~Minkowiski, Phys.\ Lett.\ B {\bf 67}, 421 (1977);
M. Gell-Mann, P. Ramond and R. Slansky in {\it Supergravity}, eds.
D. Freedman and P. Van Niuenhuizen (North Holland, Amsterdam, 1979),
p.~315; T. Yanagida in {\it Proceedings of the Workshop on Unified
Theory and Baryon Number in the Universe}, eds. O.~Sawada and
A.~Sugamoto (KEK, Tsukuba, Japan, 1979); S.L.~Glashow, {\it 1979
Carg\`ese Lectures in Physics --- Quarks and Leptons}, eds. M.~L\'evy
{\it et al.} (Plenum, New York, 1980), p.~707. See also R.N. Mohapatra and G.
Senjanovi\'c, Phys.\ Rev.\ Lett.\ {\bf 44}, 912 (1980) and J.~Schechter and J.W.F.~Valle,
   Phys.\ Rev.\  D {\bf 22}, 2227 (1980).

\bibitem{lowseesaw}
A very low-energy seesaw was proposed in A.~de Gouv\^ea,
  Phys.\ Rev.\  D {\bf 72}, 033005 (2005). See also
  A.~de Gouv\^ea, J.~Jenkins and N.~Vasudevan,
  Phys.\ Rev.\  D {\bf 75}, 013003 (2007);
F.L.~Bezrukov and M.~Shaposhnikov,
  Phys.\ Rev.\  D {\bf  75}, 053005 (2007);
  A.~de Gouv\^ea,
  arXiv:0706.1732 [hep-ph] 
for recent discussions of the phenomenology of low-energy versions of the seesaw mechanism.

\bibitem{dm_seesaw}
For a connection between dark matter and the seesaw mechanism see T.~Asaka, S.~Blanchet and M.~Shaposhnikov,
  Phys.\ Lett.\  B {\bf 631}, 151 (2005);
T.~Asaka and M.~Shaposhnikov,
  Phys.\ Lett.\  B {\bf 620}, 17 (2005);


\bibitem{deGouvea:2008nm}
  A.~de Gouv\^ea and J.~Jenkins,
  Phys.\ Rev.\ D {\bf 78}, 053003 (2008).

\bibitem{Cirelli:2004cz}
  M.~Cirelli, G.~Marandella, A.~Strumia and F.~Vissani,
  Nucl.\ Phys.\  B {\bf 708}, 215 (2005).

\bibitem{Bandyopadhyay:2007rj}
  A.~Bandyopadhyay and S.~Choubey,
  arXiv:0707.2481 [hep-ph].

\bibitem{Huber:2004xh}
  P.~Huber and T.~Schwetz,
  Phys.\ Rev.\  D {\bf 70}, 053011 (2004).
  
\bibitem{Murayama:2000iq}
  H.~Murayama and A.~Pierce,
  Phys.\ Rev.\  D {\bf 65}, 013012 (2002).
  
\bibitem{Vogel:1999zy}
  P.~Vogel and J.~F.~Beacom,
  Phys.\ Rev.\  D {\bf 60}, 053003 (1999).

\bibitem{Apollonio:2002gd}
  M.~Apollonio {\it et al.}  [CHOOZ Collaboration],
  Eur.\ Phys.\ J.\  C {\bf 27}, 331 (2003).
  
\bibitem{Ayres:2004js}
  D.~S.~Ayres {\it et al.}  [NO$\nu$A Collaboration],
  arXiv:hep-ex/0503053.

\bibitem{Bandyopadhyay:2007kx}
  See, for example,
  A.~Bandyopadhyay {\it et al.}  [ISS Physics Working Group],
  arXiv:0710.4947 [hep-ph].

\bibitem{Amsler:2008zz}
  C.~Amsler {\it et al.}  [Particle Data Group],
  Phys.\ Lett.\  B {\bf 667} (2008) 1.
  
\bibitem{Adamson:2008zt}
  P.~Adamson {\it et al.}  [MINOS Collaboration],
  arXiv:0806.2237 [hep-ex].
  
\bibitem{Maltoni:2007zf}
  M.~Maltoni and T.~Schwetz,
  Phys.\ Rev.\  D {\bf 76}, 093005 (2007).
  
\bibitem{Donini:2007yf}
  A.~Donini, M.~Maltoni, D.~Meloni, P.~Migliozzi and F.~Terranova,
  JHEP {\bf 0712}, 013 (2007).
  
\bibitem{Dighe:2007uf}
  A.~Dighe and S.~Ray,
  Phys.\ Rev.\  D {\bf 76}, 113001 (2007).
  
\bibitem{Goswami:2008mi}
  S.~Goswami and T.~Ota,
  Phys.\ Rev.\  D {\bf 78}, 033012 (2008).
  
\bibitem{Choubey:2007ji}
  S.~Choubey,
  JHEP {\bf 0712}, 014 (2007).

\bibitem{Conrad:2004gw}
 The large data samples which will be collected at near detectors could also allow for a new precision measurement of $\sin^2\theta_W$ via $\bar{\nu}+e$ elastic scattering, as discussed in 
  J.~M.~Conrad, J.~M.~Link and M.~H.~Shaevitz,
  Phys.\ Rev.\  D {\bf 71}, 073013 (2005).

 \end{thebibliography}
 \end{document}